\documentclass[journal]{IEEEtran}
\usepackage{multirow} 
\usepackage{xcolor,soul,framed} 
\usepackage{epstopdf}
\colorlet{shadecolor}{yellow}
\usepackage{graphicx}  
\usepackage{epsfig}
\usepackage[cmex10]{amsmath}
\usepackage{booktabs}
\usepackage{array}
\usepackage{mdwmath}
\usepackage{mdwtab}
\usepackage{eqparbox}
\usepackage{url}
\usepackage{amsmath,amsthm,amssymb}
\usepackage{hyperref}
\usepackage{xcolor}
\usepackage{comment}
\usepackage{wasysym}
\usepackage{multirow}
\usepackage{algorithmic}
\usepackage{array}
\usepackage{caption}
\usepackage{cite}
\usepackage{nomencl}
\usepackage{makeidx}
\usepackage{acronym}
\usepackage{stackengine}
\usepackage{svg}
\usepackage{mathtools}
\usepackage{comment}
\usepackage{makecell}

\let\labelindent\relax
\usepackage{enumitem}
\usepackage{subcaption}
\usepackage[percent]{overpic}

\newtheorem*{req}{Requirement}
\theoremstyle{definition}
\newtheorem{definition}{Definition}
\ifCLASSINFOpdf
\else
\fi

\begin{document}
\bstctlcite{IEEEexample:BSTcontrol}
    \title{Definition and Formulation of Inertia Service Incorporating Inverter-Based Resources}
  \author{Sojin Park,~\IEEEmembership{Student Member,~IEEE},~Ross Baldick,~\IEEEmembership{Fellow,~IEEE},~Hunyoung Shin~\IEEEmembership{Member,~IEEE}

    \thanks{Sojin Park and Hunyoung Shin are with the Department of Electronic and Electrical Engineering, Hongik University, Seoul 04066, Korea (e-mail: so1220os24@gmail.com; hyshin@hongik.ac.kr).}
    \thanks{Ross Baldick is with the Department of Electrical and Computer Engineering, University of Texas at Austin, Austin, TX 78751 USA (e-mail: baldick@ece.utexas.edu).}

}

\markboth{IEEE TRANSACTIONS ON POWER SYSTEMS
}{Roberg \MakeLowercase{\textit{et al.}}: High-Efficiency Diode and Transistor Rectifiers}

\maketitle

\begin{abstract}
Increasing concerns over the scarcity of inertia have motivated the procurement of inertia as an ancillary service (AS). Despite numerous academic and practical efforts, there remains a lack of consensus regarding the definition and treatment of inertia service in market operations, particularly the specification of inertia variables and the separation between synchronous inertia (SI) from synchronous generators and virtual inertia (VI) from inverter-based resources. To address these issues, this paper proposes a power-oriented (P-oriented) definition based on inertial response, which establishes conceptual consistency between SI and VI and makes the inertia service commensurable with other ASs. This definition explicitly incorporates both positive and negative inertial responses during frequency drop events. We then formulate a security-constrained economic dispatch framework based on this P-oriented definition and demonstrate its practical effectiveness through simulations. Case studies on a modified IEEE 30-bus system show that the proposed bidirectional service definition ensures price signals that reflect the economic value of inertial provision. 

\end{abstract}



%
\IEEEpeerreviewmaketitle

\vspace{-4mm}


\section{Introduction}\label{Intro}
\vspace{-0.5mm}
\IEEEPARstart{I}{n} recent years, inverter-based resources (IBRs) such as renewables and energy storage systems (ESSs) have increasingly replaced synchronous generators (SGs), reducing system inertia which is the rotational kinetic energy (KE) in SGs that is released instantaneously to arrest frequency declines. Consequently, during power disturbances such as generator or transmission line faults, the rapid frequency declines cannot be adequately mitigated by existing ancillary services (ASs) including primary frequency response (PFR), due to their inability to respond in time \cite{Denholm2020InertiaSpin}. As inertia becomes scarce, it acquires clear economic value, motivating the design of market mechanisms that compensate providers and ensure timely procurement.

In this context, numerous efforts have been made to incorporate inertia into market operations in both practical and academic domains. On the practical side, many Independent System Operators (ISOs) and Regional Transmission Organizations, responsible for maintaining frequency stability, have begun procuring inertia as an AS. For example, Ireland's transmission system operator has contracted SGs through the Delivering a Secure, Sustainable Electricity System program \cite{2024DS3Document}. The Australian Energy Market Operator also procures inertia through the Rate of Change of Frequency (RoCoF) Control Service in the Wholesale Electricity Market of Western Australia \cite{Ryan2023WEMQuantities}. Moreover, the Australian National Electricity Market, which covers the eastern states in Australia, has been actively discussing the design and implementation of an inertia market \cite{AEMC2025InertiaDraft}.

Academic research has also explored the integration of inertia services into market optimization problems. Initial studies on inertia \cite{Teng2016StochasticRequirements, Li2020PFRSystem, Matamala2024CostServices, Ela2014MarketDesign, Ela2014MarketStudies} primarily scheduled synchronous inertia (SI) from SGs through unit commitment (UC) with binary variables that represent on/off status. To enable pricing, some studies relaxed these binary variables by fixing commitment decisions.

\begin{table*}[htb!]
    \centering
    \caption{References}
    \renewcommand{\arraystretch}{0.9} 
    \vspace{-2mm}
    \begin{tabular}{c||>{\centering\arraybackslash}m{2.9cm}|>{\centering\arraybackslash}m{2.25cm}|c|c|c|>{\centering\arraybackslash}m{2.25cm}c}
    \toprule
         \textbf{References} & \textbf{Optimization Problems} & \textbf{Ancillary Services} &  \textbf{Including VI} & \textbf{Pricing} &  \textbf{Single/Double Inertia Products} &  \textbf{Units of Inertia}\\
    \midrule
         \cite{Teng2016StochasticRequirements} & UC &  Inertia, PFR & - & - & - & MW·s/Hz \\
         \cite{Li2020PFRSystem} & UC &  Inertia, FFR, PFR & - & - & - & MW·s \\
         \cite{Badesa2018OptimalLargest-Power-Infeed-Loss} & UC &  Inertia, PFR & - & - & - & MW·s \\
         \cite{Puschel-LvengreenFrequencySize} & ED &  Inertia, PFR & - & - & - &  MW·s/Hz  \\
         \cite{Matamala2024CostServices} & UC &  Inertia, PFR & - & \checkmark & - &  MW·s  \\
         \cite{Ela2014MarketDesign, Ela2014MarketStudies} & UC &  Inertia, PFR & - & \checkmark & -&  MVA·s \\
         \cite{Badesa2020PricingFormulation} & ED &  Inertia, FR & -  & \checkmark & - &  MW·s  \\
         \cite{Paturet2020StochasticGrids} & UC &  Inertia, PFR & \checkmark  & - & - &  s  \\
         \cite{Shen2023OptimalDispatch} & ED &  Inertia, PFR & \checkmark  & - & - &  MW·s  \\
         \cite{She2024VirtualSystems} & ED &  Inertia, PFR & \checkmark  & - & - &  MW·s  \\
         \cite{Kushwaha2023ASystems} & UC &  Inertia, PFR, Reserve & \checkmark  & - & - &  MW·s/Hz  \\
         \cite{Liang2023InertiaMarkets} & UC &  Inertia, Regulation & \checkmark  & \checkmark & Single &  MW·s  \\
         \cite{Li2023MarketMarket} & UC &  Inertia, PFR & \checkmark  & \checkmark & Single & MW·s/Hz  \\
         \cite{Li2025PricingNonconvexity} & UC &  Inertia, FFR, PFR & \checkmark  & \checkmark & Single &  MW·s/Hz  \\
         \cite{GarciaPrimaryResources, Garcia2025Co-OptimizationMarket} & UC &  Inertia, FFR, PFR & \checkmark  & \checkmark & Double &  SI: MW·s, VI: MW  \\
         \cite{Badesa2023AssigningSources} & UC &  Inertia, FFR, PFR & \checkmark  & \checkmark & Double &  SI, VI: MW·s \\
         \cite{Paturet2020EconomicSystems} & UC &  Inertia & \checkmark  & \checkmark & Single &  s \\
         \cite{Qiu2024MarketApproach} & UC &  Inertia, FFR, PFR & \checkmark  & \checkmark & Single &  MW·s \\
    \bottomrule 
    \end{tabular}
    \label{tab: ref}
    \vspace{-4mm}
\end{table*}

With advancements in control technologies for IBRs, attention has been drawn to grid-forming (GFM) converters, which can provide virtual inertia (VI) that emulates SI behavior. Recent research has focused on co-optimizing SI and VI within market frameworks \cite{Paturet2020StochasticGrids, Shen2023OptimalDispatch, Badesa2023AssigningSources, Li2023MarketMarket, Garcia2025Co-OptimizationMarket, Liang2023InertiaMarkets, She2024VirtualSystems, Li2025PricingNonconvexity, Kushwaha2023ASystems, Qiu2024MarketApproach}. For instance,  \cite{Badesa2023AssigningSources} modeled both SI and VI in units of MW·s through frequency-constrained unit commitment (FCUC) and derived two separate inertia prices reflecting VI’s recovery behavior. Another study \cite{GarciaPrimaryResources} treated VI as inertial power in MW, while modeling SI as KE, and accounted for nonlinear interactions between inertia and PFR and fast frequency response (FFR). On the other hand, other studies integrated SI and VI into a single inertia product with one inertia price. For example, \cite{She2024VirtualSystems} formulated security-constrained economic dispatch (SCED) with inertia variables in MW·s units and adjustable inertia constants for VI providers. Similar approaches were adopted in studies that incorporated VI from wind turbines \cite{Kushwaha2023ASystems} and ESSs \cite{Li2025PricingNonconvexity} into FCUC models using MW·s/Hz units.

Table~\ref{tab: ref} summarizes representative studies that incorporate inertia as a product in market optimization. Despite extensive academic efforts, the literature reveals significant inconsistencies in both conceptual approach and mathematical formulation. The units used for inertia variables span from traditional MW·s reflecting energy-based perspectives, to MW representing power-based approaches, and even MW·s/Hz and seconds. Moreover, some studies implement separated inertia products treating SI and VI as distinct services, while others advocate for a unified single product. This lack of consensus underscores the need for a clear and standardized definition of inertia service. Since the introduction of new service variables in market optimization problems is directly linked to valuation and compensation mechanisms, conceptual agreement on service definition must be established before successful inertia market implementation can be achieved.

This observation also raises a deeper conceptual question that has received limited attention in the literature. Although the traditional concept of inertia, rooted in the stored KE of SGs, has served the power industry well for decades, it is not clear that this definition remains the most appropriate conceptual framework for market design even with increasing penetration of IBRs. While it is intuitive to maintain consistency with the established convention when incorporating VI into market operations, we tend to overlook what aspects of SG behavior VI truly emulates, and the essence of what we call \textit{inertia}. 

Answering the questions, in this paper we propose a power-oriented (P-oriented) definition of inertia service. Rather than focusing on the total stored energy, this approach centers on the inertial response that resources actually contribute to the system. This perspective enables inertia service to be quantified based on potential delivery, similar to other ASs, making it directly commensurable with existing products and facilitating integrated co-optimization. Building upon this P-oriented definition, we also develop a comprehensive SCED framework that integrates both SI and VI providers. The main contributions of our work are as follows.

\begin{itemize}[left=0pt]
    \item This work establishes a theoretical and conceptual foundation for integrating SGs and IBRs for inertia service. By shifting focus from energy-based quantification to power-based inertial response, we provide the P-oriented perspective that helps conceptually integrate SI and VI based on their equivalent contributions to the grid, and market-oriented justification for treating SI and VI under a unified service product.
    \item Recognizing the bidirectional nature of the inertial response itself, we design a bidirectional inertia service. We explicitly include the recovery phase---often overlooked in prior market studies---into the service definition, highlighting the functions of negative response: to prevent rapid frequency surges during recovery phase and to restore energy state in a stable manner. This helps mitigate potential disputes over energy reservation responsibilities between ISOs and VI providers, guaranteeing continuous service deployment throughout the dispatch interval, even in the case of consecutive frequency events.
    \item Based on the proposed definition, we formulate a comprehensive SCED model that co-optimizes energy, inertia, and PFR. Our model includes detailed individual constraints for both power and energy aspects of inertia provision, frequency constraints, and treating the largest contingency size as a variable to provide modeling flexibility.  The effectiveness of the P-oriented formulation is demonstrated through simulations conducted on an IEEE test system.
\end{itemize}
The remainder of this paper is organized as follows. Section II presents the theoretical background on the two types of inertial responses and non-discriminatory pricing of them. In Section III, we introduce our P-oriented definition and the requirements for the inertia service. Section IV describes the formulation of constraints for SCED, and Section V presents the whole SCED formulation and derives service prices and settlements. Section VI demonstrates the effectiveness of the proposed definition and formulations through simulations. Finally, Section VII concludes the paper.
\vspace{-1mm}
\section{Background}\label{background}
\vspace{-0.5mm}
\subsection{Inertial Response}\label{subsec: Inertial Response}\vspace{-1mm}
Inertia, an inherent property of rotating masses resisting changes in rotational speed, is predominantly supplied by online SGs in power systems. For each SG $k$, an imbalance between mechanical input power (\!$\Delta P_{{\rm m},k}$\!) and electrical output power (\!$\Delta P_{{\rm e},k}$\!) causes the SG to exchange KE with the grid:
\vspace{-1mm}
\begin{align}
    \Delta P_{{\rm m},k} - \Delta P_{{\rm e},k} = \frac{d
    E_{k}}{dt}, \label{eqn: dE/dt}
\end{align}
where $E_{k}$ is the KE stored in the rotating mass of generator $k$. Assuming that rotor speed is tightly regulated around its rated value, $\omega_k = \omega_{k0}$, the KE of generator $k$, $E_{k}=\frac{1}{2}J_{k}\omega_{k}^2$, can be expressed as $H_{k}S_{k}$, where $J_{k}$ is the moment of inertia of generator $k$, $\omega_{k}$ is the angular speed, $H_{k}$ is the inertia constant in seconds, and $S_{k}$ is the MVA rating of generator $k$. Eq. (\ref{eqn: dE/dt}) can be expressed as the \textit{swing equation} with the speed in Hz:
\vspace{-3mm}
\begin{align}
 \Delta P_{{\rm m},k} - \Delta P_{{\rm e},k} = 2\frac{H_{k}S_{k}}{f_{k0}}\Delta f_k', \label{eqn: swing eqn}
\end{align}
where superscript $'$ means derivative with respect to time.
This equation describes how the generator's speed\textemdash and thus the system frequency\textemdash changes in response to power imbalance \cite{PrabhaKundur1994PowerControl}. For system-wide analysis, individual generator swing equations can be aggregated under the assumption that generators are coherently swinging, allowing the representation of system inertia as the sum of the KE of individual generators. Therefore, Eqs. (\ref{eqn: dE/dt}) and (\ref{eqn: swing eqn}) imply that change in the frequency is resisted by variations in the KE of the SG rotors, which manifests as deviations in their instantaneous power output, in proportion to minus the RoCoF, i.e., $-\Delta f'$. This inertial response occurs naturally in the direction opposite the frequency change and is provided consistently, regardless of the sign of the frequency deviation (FD) or RoCoF \cite{Su2020FastCharacteristics}.

High system inertia is crucial for frequency stability, as it mitigates the initial RoCoF following a contingency. This slower frequency decline provides the necessary time for frequency response resources to act before frequency reaches critical levels, such as Under-Frequency Load Shedding (UFLS) triggers or RoCoF protection relay limits \cite{Electric2023WhatOffer}. Consequently, the permissible RoCoF becomes an explicit operational constraint.

\vspace{-4.5mm}

\subsection{The Future Provision of Inertial Response} \label{subsec: Future Provision}
\vspace{-0.5mm}
\subsubsection{Inertia as a Service} \label{subsubsec: Service}
Traditional electricity markets have not defined inertia as an independent product nor explicitly compensated its provision because system inertia, primarily from SGs, was abundant and its marginal production cost was effectively zero. However, as SGs retire and are replaced by IBR lacking \textit{physical} inertia, there will eventually be insufficient inertia in the system to maintain frequency stability unless its provision is either remunerated through market mechanisms or mandated by regulation. To attract inertia provision, markets should therefore recognize and compensate inertia as a service. This not only contributes to securing physical inertia, but also sends efficient investment signals for VI technologies and the retention of synchronous capacity. 

In fact, inertia holds intrinsic economic value: higher inertia reduces demand for rapid frequency control ancillary services (FCASs) (e.g., PFR and FFR), while lower inertia increases it. This complementarity justifies remunerating inertia. 

\subsubsection{Non-discriminatory Compensation for SI and VI} \label{subsubsec: Nondiscriminatory}

As a general principle, all resources capable of providing inertia should be eligible for equivalent compensation, assuming they contribute equally to the grid\footnote{Although VI resources may exhibit some time delay, if the delay remains within a tolerable range relative to the time scale of the frequency event, VI can effectively substitute for SI. Otherwise, a minimum level of SI may be required.}. However, SI is an inherent property of SGs that does not incur any operating costs, so it is sometimes argued that explicit economic compensation for inertial response is unnecessary for these resources. 

Consider a market that remunerates only VI providers while no compensation is provided for SI. The Revenue Equivalence Theorem \cite{Klemperer1999AuctionLiterature} suggests that SGs would attempt to adjust their offer behavior in other markets, such as for energy, to recover the revenue they would have earned from explicit inertia payments, tending toward a total revenue similar to that under explicit inertia compensation. However, such strategic offer adjustments are unlikely to fully offset the lack of SI remuneration in complex electricity markets \cite{Baldick2009SingleMarkets}. Thus, SGs are unable to recover the full value of an explicit inertia payment, and their profitability will be diminished compared to a scenario with direct compensation. This sends a biased economic signal that potentially encourages the premature retirement of SGs and undermines efficient inertia management.

Therefore, our work defines an inertia service incorporating both types of responses. To ensure eligible providers for the inertia service are subject to non-discriminatory pricing alongside SGs, it is assumed throughout the paper that VI providers refer specifically to those capable of delivering inertial responses comparable to SI.

\subsubsection{Characteristics of IBRs for the Provision of VI} \label{subsubsec:  Virtual Inertia}

Unlike SI, the power contribution from a VI resource is strictly limited by the nominal power capacity of the inverter. To provide VI, an IBR typically must operate below its rated output to reserve headroom, and in some cases, footroom for power absorption during frequency rises. Moreover, since an IBR lacks physical rotating mass, it requires an accessible energy source to emulate the energy delivery of SI. This requires not only available energy storage capacity\footnote{Throughout the paper, we assume that an IBR has storage to provide ASs.} (e.g., batteries or capacitors) but also sufficient reserved energy specifically for providing VI response. 

Prior to significant IBR penetration, AS market definitions emphasized power capacity, as SGs faced few short-term energy constraints. Consequently, market designs typically gave little consideration to the required duration or energy for AS deployment\footnote{There have been some exceptions. For instance, spinning reserve requirements often include maximum deployment durations (e.g., 15, 30, or 60 minutes). However, these duration limits are generally related to ensuring that capacity is available again for potential subsequent contingencies, rather than fuel or energy availability constraints.}. 
For IBRs, however, ensuring the deployment of service for an indefinite duration or in response to multiple events is often challenging due to their limited energy capacity. Instead, the energy requirements associated with AS provision by IBRs generally need to be explicitly defined in advance. This poses a significant challenge for system operators and resource owners, as predicting the timing and characteristics of contingencies is technically complex. Furthermore, because the opportunity cost of AS provision varies with the reserved energy, IBR owners are highly sensitive to its accurate calculation and specification. 

In the case of inertia, this concern regarding the energy requirement can be significantly alleviated through an appropriate definition of the inertia service. As will be presented in the next section, a carefully constructed definition of inertia service can explicitly quantify an associated energy requirement for its provision, ensuring the continuous deployment of service within a committed dispatch interval.
\vspace{-1mm}
\section{Developing a Definition for Inertia Service}\label{sec: Def}
\vspace{-0.5mm}
This study aims to rigorously define inertia as an AS procured through the spot market and to develop a real-time market mechanism that incorporates it. Through formal definition and general formulations, this work seeks to ensure that both SI and VI providers are adequately compensated for their participation, while maintaining frequency stability in future low-inertia power systems.

\vspace{-2.5mm}
\subsection{Motivation for a P-oriented Perspective of Inertia Service} \label{subsec: P-oriented Perspective}
\vspace{-0.5mm}

The motivation for a new perspective stems from two critical limitations in the KE-based definition of inertia service. First, inertia in MW$\cdot$s essentially represents the hypothetical total energy deliverable to the grid if system frequency fell from nominal to zero. However, only a small fraction of the total stored KE is practically utilized, as system frequency must be tightly regulated near its nominal value. Second, the concept of physical KE is not directly applicable to IBRs, for which inertial response is not an inherent physical property but a function of control algorithms. This disconnect poses a fundamental challenge for integrating IBRs into inertia provision on comparable terms with SGs.

These limitations motivate a shift toward a \textit{P-oriented perspective}. This perspective defines the inertia service based on the active power (MW) a resource provides in proportion to RoCoF. This approach effectively resolves the aforementioned challenges. By defining the service in terms of power delivery, it establishes the necessary commensurability with other AS products. Furthermore, this power-based definition is technologically neutral, allowing both SGs and IBRs to be evaluated and compensated based on their actual performance. Consequently, it enables the consistent integration of inertia into established, co-optimized market mechanisms for modern power systems with high IBR penetration.

\begin{figure}
  \centering
  \includegraphics[width=2.6in, height=1.7in]{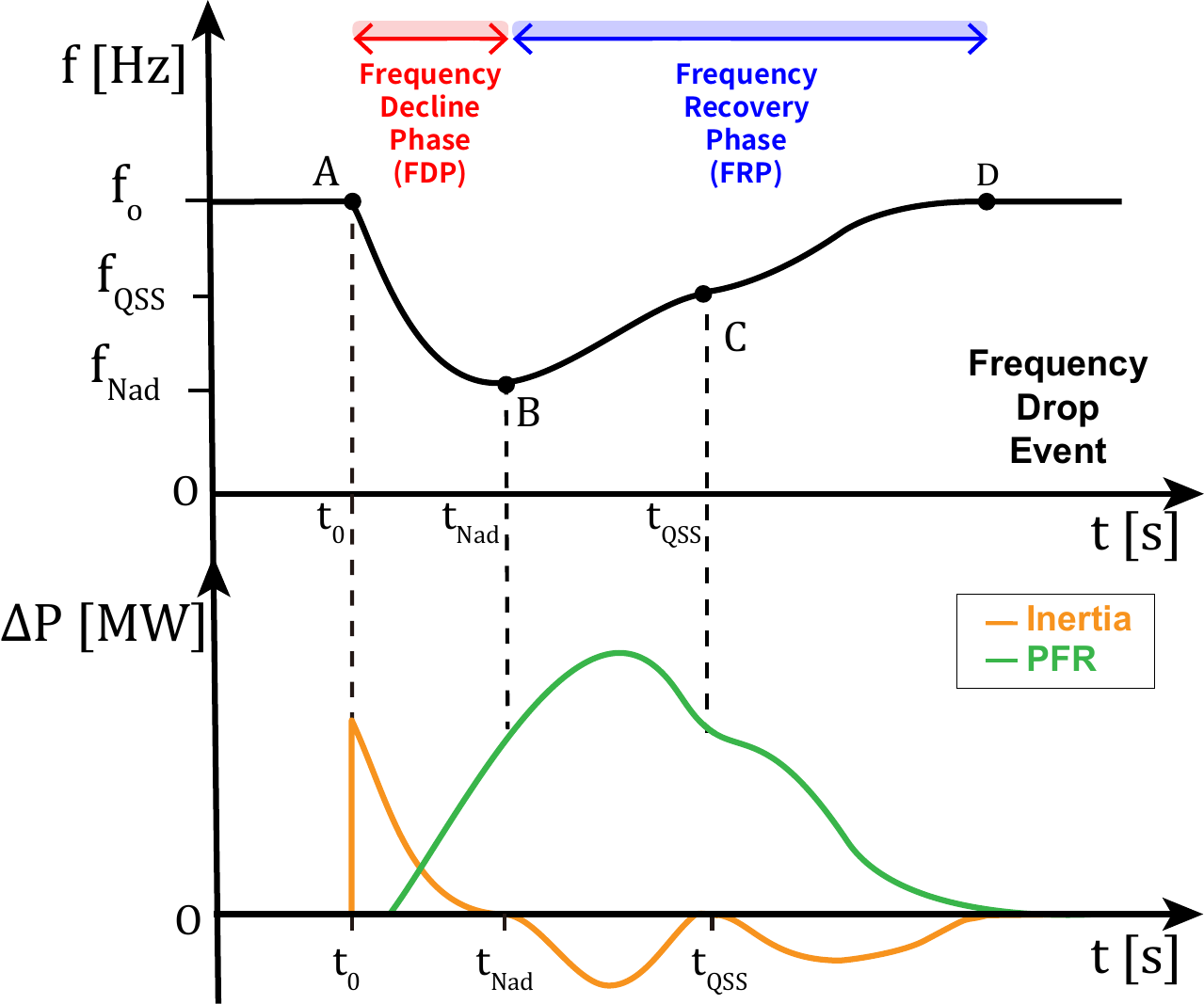}
  \vspace{-1.5mm}
  \caption{Frequency trajectory and responses after contingency}
  \vspace{-6mm}
  \label{fig: frequency drop}
\end{figure}
\vspace{-3mm}

\subsection{A Definition of Inertia Service} \label{subsec: Definition}
\begin{definition}
\vspace{-0.5mm}
The \textit{inertia provision} for a provider $k$ is the change in power to resist RoCoF, represented as: 
\vspace{-1mm}
\begin{align}
\Delta P^{\rm IP}_k\coloneq - D_k \Delta f'.\label{eqn: Def}
\end{align}
where $D_k$ denotes the proportionality factor [MW/(Hz/s)] that links the change in power to RoCoF for each resource. 
\vspace{-1mm}    
\end{definition}

For SGs, $D_k$ is a fixed mechanical property, whereas for VI providers, it is a tunable control parameter. Although similar definitions have been adopted in previous studies (e.g., \cite{GarciaPrimaryResources, Badesa2023AssigningSources, Garcia2022RequirementsControl}) for VI products, this study is the first to define inertia service in MW units for SI, as well as for VI, within a unified market framework.

This definition reflects the bidirectional nature of inertial response. As illustrated in Fig.~\ref{fig: frequency drop}, inertia not only provides a positive response to arrest the frequency decline during the A--B interval but also contributes a negative response during the B--D recovery interval to suppress excessive frequency rebound. Accordingly, the inertia provision encompasses not only the immediate post-fault action during the frequency decline phase (FDP) but also the subsequent response during the frequency recovery phase (FRP).

To provide FCASs, a resource must commit a predefined quantity of power as operating reserve to the system operator and retain the capability to deliver, upon a contingency, a response up to the committed quantity. For the inertia service, while SGs can supply a response that is effectively unconstrained by nameplate capacity, IBRs are limited by inverter physical power capability (e.g., semiconductor current limits). Hence, the quantity of service must be clearly specified for IBRs. Importantly, to reflect the bidirectional nature of inertial response, this quantity must be specified for both headroom (positive response) and footroom (negative response).  This leads to the formal definition of the inertia service:
\vspace{-1mm}
\begin{definition}
Given the maximum allowable magnitude of RoCoF ($\overline{\Delta f'}$), the \textit{inertia service} is the commitment of a provider $k$ to supply an \textit{inertia provision} in a bidirectional way following a frequency drop event, with the committed service quantity of $P_k^{\rm In} := D_k\overline{\Delta f'}$ in each direction.  
\end{definition}
\vspace{-1.5mm}
\begin{figure}
  \centering
  \includegraphics[width=2.5in, height=0.9in]{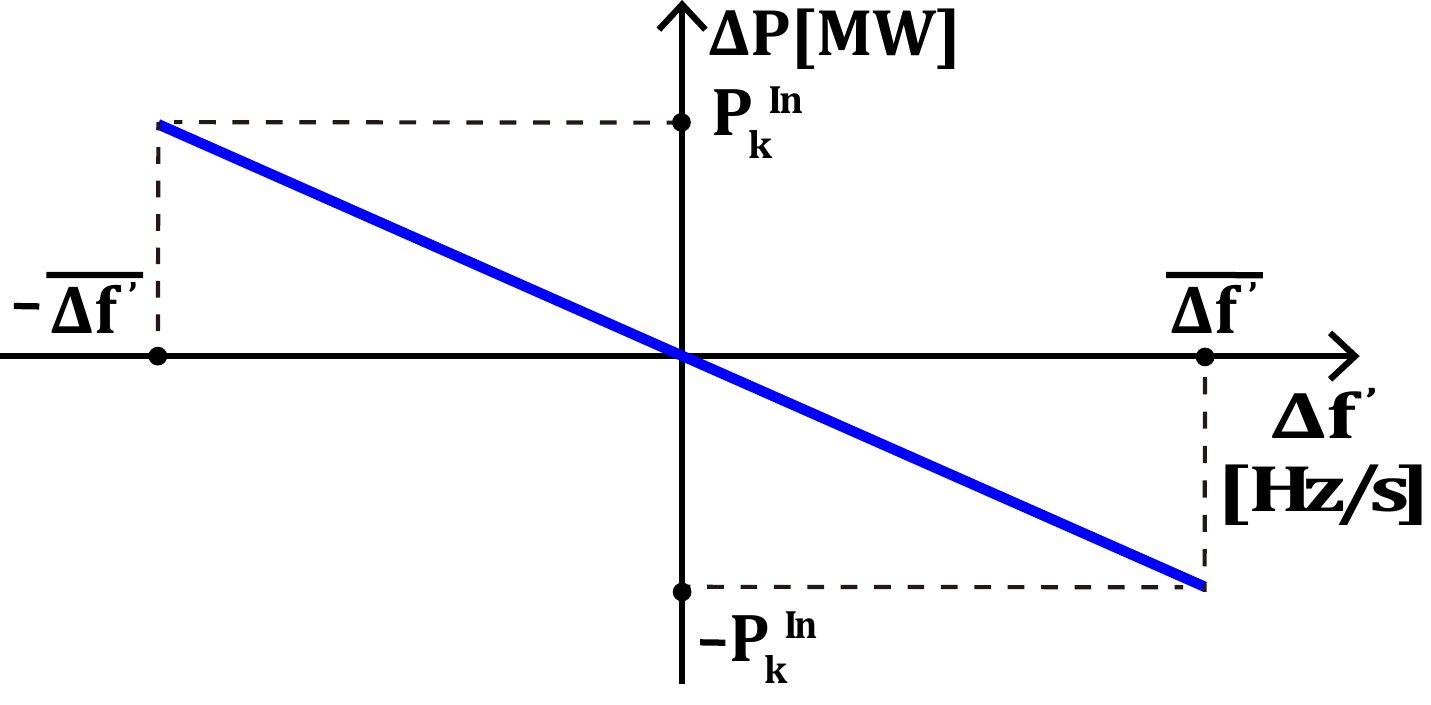}
  \vspace{-1.5mm}
  \caption{P-oriented definition of inertia service}
  \label{fig: Def}
  \vspace{-7mm}
\end{figure}

The symmetrical quantity is specified because although the RoCoF magnitude during the FRP is typically smaller than its initial value, its unpredictability makes this conservative approach prudent for ensuring secure system operation. This requirement may be relaxed in practice depending on operational conditions. Fig.~\ref{fig: Def} illustrates the inertia service. For a given $\overline{\Delta f'}$, a provider supplies a bidirectional response up to its committed service quantity, following a slope determined by its characteristics\footnote{Although the curve may resemble a droop curve, it differs in two respects: the output is an instantaneous response proportional to minus the RoCoF, and for IBRs the slope can be adjusted in each dispatch interval through its control scheme, as part of a market strategy.}. 

The P-oriented definition allows for the direct relationship between inertia service quantity and system security: the sum of the inertia service quantities from all providers must be at least as large as the largest contingency. This would ensure that, in the event of the contingency, the RoCoF would not exceed the maximum allowable magnitude of RoCoF (See constraint (\ref{eqn: RoCoF}) in Section \ref{subsec: Frequency constraints}).

As noted in Section \ref{subsubsec:  Virtual Inertia}), the energy requirement for the inertia service must be specified. The committed service must be repeatedly deployable during successive contingencies, independent of frequency trajectory, provided the FD remains within the maximum allowable frequency deviation (MAFD).
\vspace{-4mm}
\begin{req}
Inertia providers must reserve footroom in their energy capacity to inject the maximum energy contribution corresponding to the MAFD at the frequency nadir. The maximum energy contribution of a provider $k$ is defined as follows:
\vspace{-3mm}
\begin{align}
E_{k}^{\rm In} = {D_k} \overline{\Delta f}_{\rm Nad} = \frac{\overline{\Delta f}_{\rm Nad}}{\overline{\Delta f^{'}}} P_k^{\rm In}.\label{eqn: Ereq}
\end{align}
\end{req}
\vspace{-2mm}
Because the provision of inertial energy is proportional to the FD, the energy released during the FDP is, in principle, fully recovered during the FRP. For an IBR, however, a practical energy reservation must account for round-trip efficiency to compensate for any net energy depletion over the event. That is,  Fig.~\ref{fig: Def} should be slightly modified to have a steeper slope in the case of positive RoCoF.
\vspace{-4mm}
\subsection{Benefits of the P-oriented Service Definition} \label{subsec: Benefits}
\vspace{-0.5mm}

\subsubsection{Enabling Market Co-optimization through a Commensurable Definition} \label{subsubsec: Benefit co-optimization}

The primary advantage of the proposed P-oriented definition is that it makes the inertia service commensurable with power production and other ASs. By expressing the inertia service in MW with an associated maximum power contribution, the service integrates directly into existing co-optimization market frameworks. Co-optimization fundamentally requires all products to be expressed in a common power unit so that the market can evaluate opportunity costs against each resource's capacity. In contrast, defining inertia only by an inertia constant or stored KE, without a corresponding maximum power quantity, leaves the service incommensurable with other power-based services and impedes seamless co-optimization.

This commensurability allows the full economic value of the inertia service to be captured through existing compensation mechanisms, which use co-optimization to price the opportunity cost of reserved capacity. One key feature of this valuation is the explicit inclusion of the negative response (footroom). While less intuitive, inertia inherently resists frequency changes in both directions. The proposed framework properly values this bidirectional contribution, ensuring that economic compensation reflects the full service provided.
\subsubsection{Clarifying Service Requirements for Reliable Provision} \label{subsubsec: Benefit clarification}

The service requirements must be clear and robust, particularly for IBRs. The P-oriented definition achieves this by deriving explicit power and energy commitments directly from established system security limits, such as the maximum allowable RoCoF and MAFD. This gives providers a clear standard for reserving the required power and energy capacity, thereby removing a significant barrier to market participation.

Furthermore, this definition ensures reliable and continuous service delivery. Sustained deployment across successive contingencies requires energy recovery during the FRP. For SGs, the rotor naturally reabsorbs energy, which is economically equivalent to purchasing energy at real-time price (RTP). To provide a comparable level of reliability, an IBR must also restore its state of charge (SoC), necessitating energy purchases at the RTP. By defining the negative response as an intrinsic part of the service, our approach guarantees this necessary energy recovery. This eliminates the risk of ad-hoc recovery procedures triggering system instability events such as second frequency dip. This clarification of operational requirements thus benefits both system operators by ensuring reliable service and market participants by providing clear energy requirements.

\subsubsection{Ensuring Bidirectional Coverage Through Up and Down Inertia Service} \label{subsubsec: Benefit up and down service}

\begin{figure}
    \centering
    \includegraphics[width=2.75in, height=1.2in]{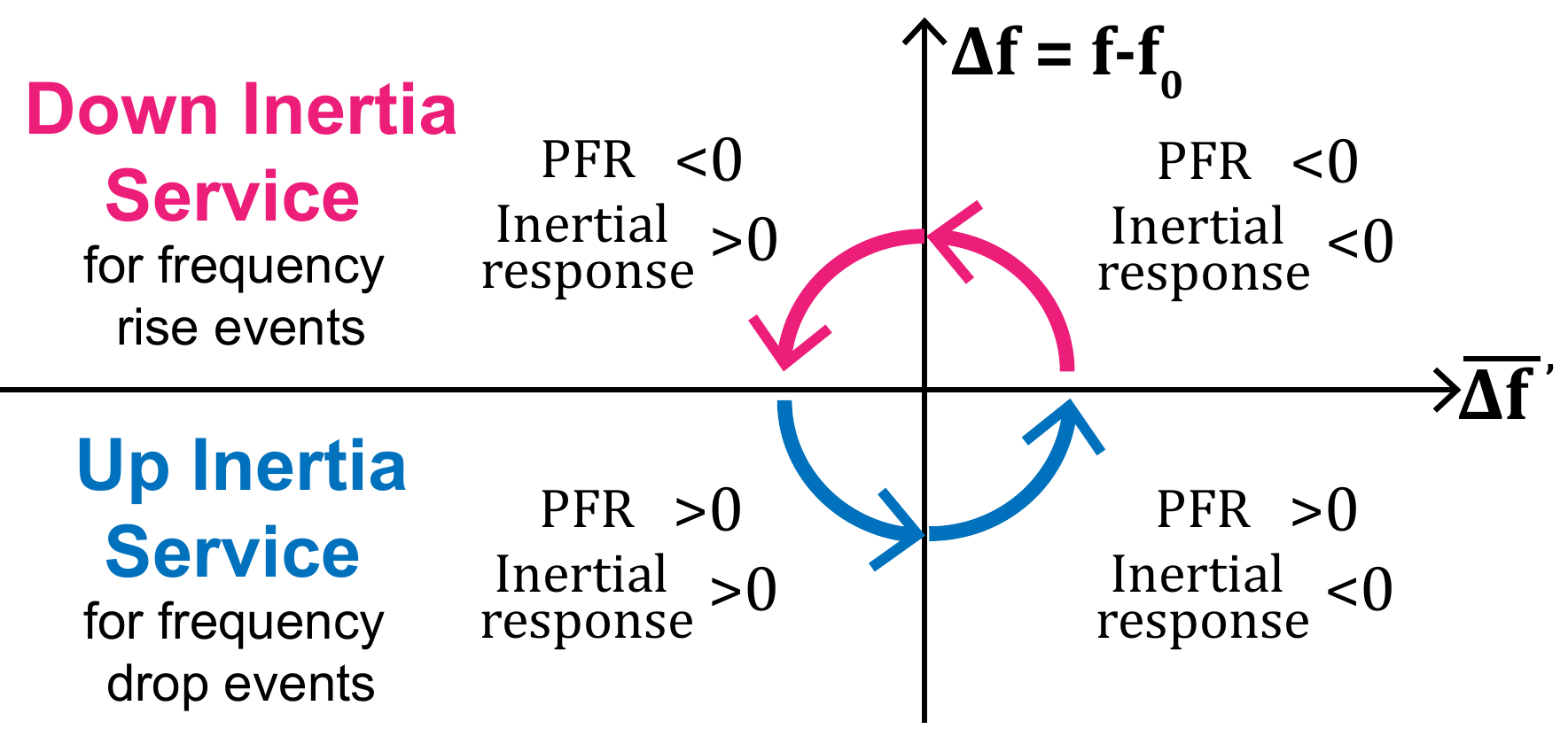}
    \vspace{-0.5mm}
    \caption{Bidirectional inertia service coverage}
    \vspace{-3mm}
    \label{fig: signs}
\end{figure}

The proposed definition extends naturally to over-frequency situations, so frequency rise events can be handled within the same bidirectional service framework used for frequency drops. Fig.~\ref{fig: signs} illustrates the signs of PFR and inertial response depending on the signs of FD and RoCoF ($\Delta f'$). The region with $\Delta f <0$ defines the operating domain for \textit{Up inertia service} (frequency drop events) and the region with $\Delta f > 0$ defines the domain for \textit{Down inertia service} (frequency rise events). 

For the down inertia service, the required power and energy commitments can be derived from system security limits for over-frequency, such as a maximum positive RoCoF and a MAFD at the frequency zenith. The operational constraints used in market operations for this service can be straightforwardly derived by reversing the signs of the up inertia service. 

This ultimately leads to a unified \textit{Up and Down inertia service}. In such a service, the required power capacities would be the maximum of the up or down requirements in each direction, and the energy capacity would be a bidirectional reserve. The practical importance of this unified capability is highlighted by incidents like the 2021 Chilean blackout \cite{Chile}, which involved both significant frequency drops and rises. 
\vspace{-3mm}
\section{Developing Constraint Formulation}\label{sec: Cons}
\vspace{-0.5mm}
This section derives constraint formulations for co-optimizing energy, inertia, and PFR, based on the inertia service definition from Section~\ref{sec: Def}. We begin by adding a model of PFR, then formulate individual resource constraints for power and energy capacity. Subsequently, we introduce system-wide frequency security constraints derived from representative limits and, finally, discuss the modeling of the largest contingency as a decision variable.
\vspace{-4.25mm}
\subsection{PFR Provision} \label{subsec: PFR Provision}
\vspace{-0.5mm}
In this study, PFR is defined as an AS in which a resource provides a steady-state power output proportional to the FD. The magnitude of this response is determined by the resource's droop characteristic, denoted as $R$. 

To effectively co-optimize inertia and PFR, it is crucial to quantify their respective contributions to maintaining frequency security limits, particularly at the frequency nadir. The PFR contribution at the nadir, however, is determined by its transient response, not its quasi-steady-state (QSS) output. Accurately predicting this transient behavior is  challenging, as it depends on system operating conditions and the types of contingencies. To address this modeling challenge, we employ a two-stage linear approximation of PFR provision. The initial PFR delivery up to its peak is modeled as a  ramp, while the subsequent output is modeled as a constant response corresponding to the QSS. This approximation enables tractable optimization while capturing the essential dynamics of PFR. 

We denote the decision variables representing power capacity reservation for the ramp response and droop response for a resource $k$ as $P_k^{\rm PFR,r}$ and $P_k^{\rm PFR,d}$, respectively. Assuming that the ramp response must be fully activated by time, $T^{\rm PFR}$, as specified by the system operator, the PFR provision from resource $k$ can be expressed as:
\vspace{-1mm}
\begin{equation}
PFR_k (t) = \begin{cases}
\frac{P_k^{\rm PFR,r}}{T^{\rm PFR}}t, & {\text{if}}\ 0 \leq t \leq T^{\rm PFR} \\
P_k^{\rm PFR,d}, & {\text{if}}\ t > T^{\rm PFR},
\end{cases}
\label{eqn: PFR}
\end{equation}
which is a modification of the PFR model in \cite{Badesa2023AssigningSources}. Then the total PFR provision in the system is obtained as $PFR(t)\!=\!\sum_{k\in\mathcal{K}}PFR_k(t)$. These decision variables are limited by the maximum PFR capability of each resource, which is determined by the resource’s physical characteristics and system security limits. The maximum capacities for the ramp response, $\overline{P}_k^{\rm PFR,r}$, and for the droop response, $\overline{P}_k^{\rm PFR,d}$, are calculated using the resource's droop characteristic ($R_k$) and the system-defined MAFD at the nadir ($\overline{\Delta f}_{\rm Nad}$) and at QSS ($\overline{\Delta f}_{\rm QSS}$), respectively: 
\vspace{-1mm}
\begin{align*}
    \overline{P}_k^{\rm PFR,r} &= \frac{R_{k}\overline{P}_{k}}{f_0}\overline{\Delta f}_{\rm Nad},\quad \overline{P}_k^{\rm PFR,d} = \frac{R_{k} \overline{P}_{k}}{f_0}\overline{\Delta f}_{\rm QSS},
\end{align*}
where $\overline{P}_k$ represents the rated power of resource $k$. Accordingly, the committed PFR quantities are governed by the following constraints:
\vspace{-1.5mm}
\begin{align}
    P_k^{\rm PFR,r} &\leq \overline{P}_k^{\rm PFR,r} \label{eqn: PFRr}\\
    P_k^{\rm PFR,d} &\leq \overline{P}_k^{\rm PFR,d} \label{eqn: PFRd}\\
    P_k^{\rm PFR,d} &\leq {P}_k^{\rm PFR,r} \label{eqn: PFRrd}.
\end{align}
\vspace{-0.5mm}
Constraint \eqref{eqn: PFRrd} ensures that the droop response capacity does not exceed the ramp response capacity, reflecting the physical limitation that sustained response cannot be greater than the peak of transient response.

Unlike the inertia service, where the required energy is clearly determined by the frequency stability criteria, calculating the required energy for PFR is less straightforward. We therefore adopt a conservative, worst-case assumption: a contingency occurs at the beginning of the dispatch interval, and the PFR provision is sustained throughout the interval. Under this scenario, the required energy for resource $k$ is: 
\vspace{-1.5mm}
\begin{align}
E_k^{\rm PFR} = \frac{1}{2}T^{\rm PFR}P_k^{\rm PFR,r}+ P_k^{\rm PFR,d}(\Delta T - T^{\rm PFR}), \label{eqn: E PFR}
\end{align}
where the first term represents the energy for the ramp-up phase, and the second term represents the energy for the sustained droop response during the remaining dispatch interval. $T^{\rm PFR}$ represents the activation time, and $\Delta T$ denotes the dispatch interval.
\vspace{-5mm}

\subsection{Individual Resource Constraints} \label{subsec: indi constraints}
\vspace{-1mm}
\begin{figure}
  \includegraphics[width=3.6in, height = 1.1in]{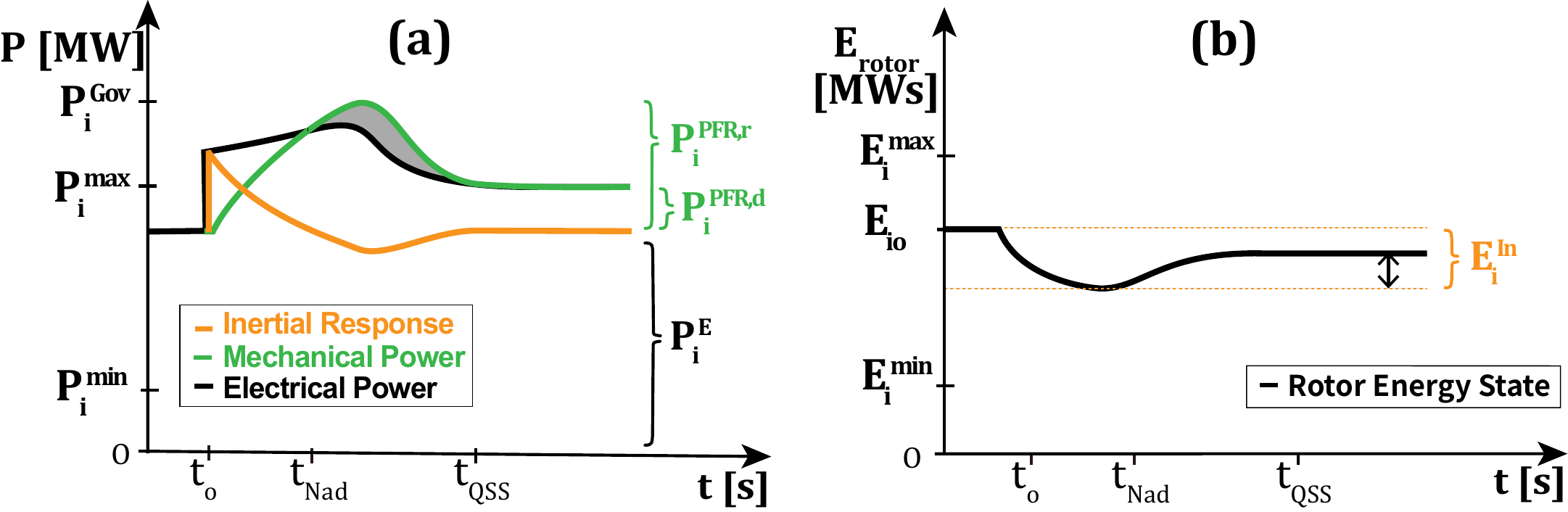}
  \caption{Synchronous generator dispatch results showing (a) power capacity allocation and (b) rotor energy dynamics}
  \label{fig: SG dispatch}
\end{figure}
\subsubsection{Power Constraints}\label{subsubsec: Power}
Power capacity constraints must ensure that a resource's total output, comprising its energy dispatch and all committed ASs, remains within its physical limits throughout the dispatch interval. The formulation of these constraints differs significantly between SGs and IBRs due to their distinct physical characteristics.

For an SG $i\in\mathcal{I}$, the inertia service does not require a dedicated power capacity reservation. Its inertial response is an inherent property, and the associated power limits are effectively non-binding. In contrast, its mechanical power limit during transient conditions may apply. This mechanical power limit, constrained by the maximum ramp response of PFR, is referred to as the governor-controlled capacity ($\overline{P}^{\rm Gov}_{i}$). The applicable constraints are therefore:
\vspace{-0.5mm}
\begin{gather}
    \underline{P}_i\leq P_{i}^{\rm E} \leq \overline{P}_{i}\label{eqn: finalGenLim}\\
     P_{i}^{\rm E} + P_{i}^{\rm PFR,r} \leq \overline{P}_{i}^{\rm Gov} \label{eqn: finalGenASLim1}\\
     P_{i}^{\rm E} + P_{i}^{\rm PFR,d} \leq \overline{P}_{i}. \label{eqn: finalGenASLim2}
\end{gather}
\vspace{-0.5mm}
Fig.~\ref{fig: SG dispatch} illustrates possible power and energy trajectories after the event. Fig.~\ref{fig: SG dispatch}(a) depicts the power dynamics, while Fig.~\ref{fig: SG dispatch}(b) illustrates the corresponding change in the rotor's KE state. Following the contingency, the electrical power output deviates from the mechanical power input due to the inertial response. Subsequently, the mechanical power increases to provide PFR. The shaded area in Fig.~\ref{fig: SG dispatch}(a) represents the energy replenished in the rotor during the negative inertial response, which directly corresponds to the energy recovery shown by the double-headed arrow in Fig.~\ref{fig: SG dispatch}(b).

In contrast, an IBR $j\in\mathcal{J}$ must provide its combined output, including inertia provision, within its inverter power rating $\left[\underline{P}_{j},\overline{P}_{j}\right]$. The following constraints (\ref{eqn: ESSLim})--(\ref{eqn: ESSASLim4}) enforce this requirement by ensuring the feasibility of the power trajectory shown in Fig.~\ref{fig: frequency drop}:
\vspace{-1.25mm}
\begin{gather}
    \underline{P}_j \leq P_{j}^{\rm E} \leq \overline{P}_{j} \label{eqn: ESSLim}\\
    P_{j}^{\rm E} + P_{j}^{\rm In}\leq \overline{P}_{j} \label{eqn: ESSASLim1}\\
    P_{j}^{\rm E} + \alpha_{j} \left(P_{j}^{\rm PFR,r} + P_{j}^{\rm In}\right) \leq \overline{P}_{j} \label{eqn: ESSASLim2}\\
    \underline{P}_j \leq P_{j}^{\rm E} + \beta_{j} P_{j}^{\rm PFR,r} - P_{j}^{\rm In}/{\eta_j}\leq \overline{P}_{j} \label{eqn: ESSASLim3}\\
    P_{j}^{\rm E} + P_{j}^{\rm PFR,d}\leq \overline{P}_{j}. \label{eqn: ESSASLim4}
\end{gather}
\vspace{-0.75mm}
Constraint (\ref{eqn: ESSASLim1}) represents the instantaneous power limit at the moment of the contingency ($t_0$), when the inertial response is maximum, but PFR has not yet activated. Constraint (\ref{eqn: ESSASLim2}) secures the required headroom before reaching the nadir ($t_{\rm Nad}$). During the FDP, the maximum inertial response ($P_{j}^{\rm In}$) and PFR ramp ($P_{j}^{\rm PFR,r}$) do not typically occur simultaneously, making their combined peak output uncertain. The relaxation parameter $\alpha_j \in [0,1]$ is therefore introduced to account for this potential non-coincidence of peak responses. Subsequently, constraint \eqref{eqn: ESSASLim3} imposes the headroom and footroom requirements during the FRP. This is critical for accommodating the negative inertial response needed for energy recovery, scaled by the round-trip efficiency ($\eta_j$) of the IBR. The parameter $\beta_j\in[0,1]$ models the extent to which a positive PFR output might offset this negative requirement. These relaxation parameters can be set by the system operator or chosen by resource owners\footnote{If the system operator chooses $\alpha_{j}$ close to 1, the headroom required for positive inertial response and PFR becomes relatively large. Conversely, a low $\beta_{j}$ value imposes a conservative footroom requirement for negative inertial response, effectively neglecting PFR contributions. To simplify implementation, ISOs can also adopt unified values for all resources (e.g., $\alpha$ and $\beta$ as system-wide parameters).}. Finally, constraint \eqref{eqn: ESSASLim4} ensures the resource can deliver its sustained droop response  ($P_{j}^{\rm PFR,d}$) at the QSS, when the inertial response is negligible.

In the remainder of this paper, we assume the conservative case where $\alpha_{j}=1$ and $\beta_{j}=0$, simplifying the constraint set to the most critical headroom and footroom limits:
\vspace{-1.5mm}
\begin{gather}
    \underline{P}_j \leq P_{j}^{\rm E} \leq \overline{P}_{j} \label{eqn: finalESSLim}\\
    P_{j}^{\rm E} + P_{j}^{\rm PFR,r} + P_{j}^{\rm In} \leq \overline{P}_{j} \label{eqn: finalESSASLim1}\\
    \underline{P}_j \leq P_{j}^{\rm E} - {P_j}^{\rm In}/{\eta_j}. \label{eqn: finalESSASLim2}
\end{gather}
\vspace{-4.5mm}
\subsubsection{Energy Constraints}\label{subsubsec: Energy}
While SGs are unconstrained in energy terms for the short term operations, an IBR's SoC must be explicitly managed to guarantee its delivery of the committed services. The energy constraints for an IBR $j\in\mathcal{J}$ are therefore formulated as follows: 
\vspace{-1mm}
\begin{gather}
\underline{E}_j + (E^{\rm PFR}_j+E^{\rm In}_j)/\sqrt{\eta_j} \leq E_{j0} \label{eqn: E stored}\\
\underline{E}_j + (E^{\rm PFR}_j+E^{\rm In}_j)/\sqrt{\eta_j} \leq E_{j} \label{eqn: E storedT}\\
E_{j} = E_{j0} + E_j^{\rm E} \label{eqn: Estate}\\
\underline{E}_j \leq E_{j} \leq \overline{E}_j,\label{eqn: Elimits}
\end{gather}
where $E_{j0}$ and $E_{j}$ denote the SoC at the beginning and end of the dispatch interval, respectively. $\underline{E}_j$ and $\overline{E}_j$ represent the minimum and maximum SoC limits, respectively. 

Constraints (\ref{eqn: E stored})--(\ref{eqn: Elimits}) establish the energy reservation required for the energy dispatch and committed ASs. The first constraint guarantees the sufficient energy for the ASs at the beginning of the dispatch interval, given the initial SoC. Constraint (\ref{eqn: E storedT}) ensures that the resource is prepared to provide continuous service until the end of the dispatch interval. Constraint (\ref{eqn: Estate}) represents the SoC dynamics with the energy requirement for energy service, $E_j^{\rm E}$, which can be positive or negative due to the bidirectional operation (discharging and charging) of the resource. The last constraint imposes the energy limits on the SoC. 
\vspace{-3mm}

\subsection{Frequency Constraints} \label{subsec: Frequency constraints}
\vspace{-0.5mm}
We construct system-level constraints to meet frequency security limits. Following \cite{Matamala2024CostServices, Badesa2023AssigningSources, Badesa2020PricingFormulation}, three constraints are derived to enforce the limits for RoCoF, frequency nadir, and QSS during frequency events. Let $P^{\rm In}$, $P^{\rm PFR,r}$, and $P^{\rm PFR,d}$ represent the total system provisions of inertia, PFR ramp, and PFR droop responses, and $\Delta P^{\rm L}$ be the largest contingency size. Under the P-oriented definition, these frequency constraints are expressed as follows:\footnote{Here, the frequency nadir constraint (\ref{eqn: Nadir}) is presented in the second-order cone programming form. The proposed inertia service definition, however, is not dependent on this specific formulation and is compatible with other forms of the nadir limit.}
\vspace{-1mm}
\begin{gather}
P^{\rm In} \geq \Delta P^{\rm L}\label{eqn: RoCoF}\\
\left(\frac{P^{\rm In}}{2\overline{\Delta f^{'}}}\cdot \frac{P^{\rm PFR,r}}{T^{\rm PFR}}\right) \geq \frac {(\Delta P^{\rm L})^2}{4\overline{\Delta f}_{\rm Nad}} \label{eqn: Nadir}
\end{gather}
\begin{gather}
P^{\rm PFR,d} \geq \Delta P^{\rm L}. \label{eqn: QSS}
\end{gather}
Constraint (\ref{eqn: RoCoF}) illustrates the conceptual clarity of the P-oriented definition. Unlike energy-based formulations \cite{Matamala2024CostServices, Badesa2023AssigningSources, Badesa2020PricingFormulation}, the power-based approach directly compares the total inertia service quantity with the contingency magnitude.
\vspace{-5mm}
\subsection{Largest Contingency} \label{subsec: Largest C}
\vspace{-0.5mm}
The largest contingency size ($\Delta P^{\rm L}$) plays a crucial role in determining the required levels of ASs and, consequently, the overall cost of maintaining frequency security. In many systems, $\Delta P^{\rm L}$ is typically fixed as the rated capacity of the largest unit, or otherwise based on the largest credible contingency such as total capacity at a given plant or capacity of largest infeed to the system. While this conservative approach can improve the system security, it may lead to the over-procurement of ASs. 

Instead, treating $\Delta P^{\rm L}$ as a decision variable enables more economically efficient system operations. While an accurate frequency security model necessitates considering the loss of both energy and AS contributions from a tripped generator, we adopt a simplified approach following \cite{Badesa2018OptimalLargest-Power-Infeed-Loss, Puschel-LvengreenFrequencySize} for practicality\footnote{Consideration of energy and AS depletion implies that the resource binding one security constraint may differ from the one binding another. This may be unfavorable to system operators who typically rely on a consistent worst-case contingency criterion (e.g., N-1 or N-2). Furthermore, enforcing a single resource to determine the contingency across all frequency constraints requires binary variables, significantly increasing computational complexity.}. In this study, we define the largest contingency as the largest energy dispatch from any single resource. For this, we add the following constraint to \eqref{eqn: RoCoF}--\eqref{eqn: QSS}:
\vspace{-0.5mm}
\begin{align}
    P_{k}^{\rm E} \leq \Delta P^{\rm L}, \quad \forall k\in\mathcal{I}\cup\mathcal{J}.
\end{align}
\vspace{-8mm}

\section{Developing a Real-Time Market Optimization Problem} \label{sec: RTM}
\vspace{-0.5mm}
\subsection{Security-Constrained Economic Dispatch}\label{ED}
\vspace{-0.5mm}
This subsection presents the SCED formulation that co-optimizes the P-oriented inertia service with energy and PFR. We focus on the system-level constraints and representative individual resource constraints that demonstrate the essential features of the proposed P-oriented definition. 

First, the objective function of minimizing total operating costs is formulated as follows:
\vspace{-0.5mm}
\begin{align*}
\text{min}\sum_{i\in\mathcal{I}}\! {\rm C}_i(P_i^{\rm E},\!P_i^{\text{PFR,r}\!},\!P_i^{\text{PFR,d}}) \!+\!\sum_{j\in\mathcal{J}}\!{\rm C}_j(P_j^{\rm E},\!P_j^{\text{PFR,r}\!},\!P_j^{\text{PFR,d}\!},\!P_j^{\rm In}),
\end{align*}
where ${\rm C}_i(\cdot, \cdot, \cdot)$ and ${\rm C}_j(\cdot, \cdot, \cdot, \cdot)$ denote the cost functions of SG $i\in\mathcal{I}$ and IBR $j\in\mathcal{J}$, respectively. Notably, the cost of an SG does not include a term for inertia, as the marginal cost of providing SI is effectively zero. In contrast, the cost for an IBR includes a term for its inertia service commitment ($P_j^{\rm In}$), which is typically non-zero to reflect provision costs and opportunity costs of reserving energy capacity.

The system-wide constraints with their associated dual variables are presented in (\ref{eqn: PB})--(\ref{eqn: ED QSS}). Constraint (\ref{eqn: PB}) ensures that the total power supply meets the pre-contingency demand, while (\ref{eqn: Indefine})--(\ref{eqn: PFRddefine}) define the total system-wide procurement for each AS. The frequency security limits, which are coupled via the dispatchable largest contingency variable ($\Delta P^{\rm L}$), are enforced by (\ref{eqn: ED Largest C})--(\ref{eqn: ED QSS}). 
\vspace{-1mm}
\begin{align}
& (\lambda^{\rm E}): \sum_{i\in\mathcal{I}} P_i^{\rm E}+ \sum_{j\in\mathcal{J}} P_j^{\rm E} =  P^{\rm D} \label{eqn: PB} \\
& (\lambda^{\rm Inertia}): P^{\rm In} = \sum_{k\in\mathcal{I}\cup\mathcal{J}} P_k^{\rm In} \label{eqn:  Indefine}\\
& (\lambda^{\rm PFR,r}): P^{\rm PFR,r} = \sum_{k\in\mathcal{I}\cup\mathcal{J}} P_k^{\rm PFR,r} \label{eqn: PFRrdefine}\\
& (\lambda^{\rm PFR,d}): P^{\rm PFR,d} = \sum_{k\in\mathcal{I}\cup\mathcal{J}} P_k^{\rm PFR,d}  \label{eqn: PFRddefine}\\
& (\lambda^{k}): P_{k}^{\rm E} \leq \Delta P^{\rm L}, \quad \forall k\in\mathcal{I}\cup\mathcal{J} \label{eqn: ED Largest C}\\
& (\mu): P^{\rm In} \geq \Delta P^{\rm L} \label{eqn: ED RoCoF}\\
& (\gamma_1, \gamma_2, \gamma_3): \left(\frac{P^{\rm In}}{2\overline{\Delta f^{'}}} \cdot \frac{P^{\rm PFR,r}}{T^{\rm PFR}}\right)  \geq \frac {(\Delta P^{\rm L})^2}{4\overline{\Delta f}_{\rm Nad}} \label{eqn: ED Nadir}\\
& (\delta): P^{\rm PFR,d} \geq \Delta P^{\rm L}. \label{eqn: ED QSS}
\end{align}

For the individual resource constraints, in addition to (\ref{eqn: PFRr})-(\ref{eqn: finalGenASLim2}) and (\ref{eqn: finalESSLim})-(\ref{eqn: Elimits}), the following constraints specify the committed inertia quantity for each provider type must be added:
\vspace{-1mm}
\begin{gather}
    P_k^{\rm In} = D_k\overline{\Delta f^{'}}, \quad \forall k\in\mathcal{I}\cup\mathcal{J} \label{eqn: GenIn}\\
    0 \leq D_j \leq \overline{D}_{j}, \quad \forall j\in\mathcal{J} \label{eqn: ESSInLim}\\
    E_{j}^{\rm In} =  D_j\overline{\Delta f}_{\rm Nad}, \quad \forall j\in\mathcal{J}. \label{eqn: E In}
\end{gather}

For SGs, the proportional factor, $D_i$, is a fixed parameter, but for IBRs, $D_j$ is a decision variable bounded by their physical limitation, allowing optimization of their inertia provision. Constraint (\ref{eqn: E In}) determines the corresponding energy requirement for IBRs to provide inertia service. Finally, the SoC dynamics (\ref{eqn: Estate}) are refined to account for round-trip losses. To avoid introducing binary variables typically required to prevent simultaneous charging and discharging, we adopt the linear storage formulation proposed in\cite{Shen2023OptimalDispatch}.
\noindent
\vspace{-3mm}
\subsection{Pricing} \label{subsec: Pricing}
\vspace{-0.5mm}
The market clearing prices for energy and ASs can be derived from the Lagrangian of the SCED. For brevity, we present the portion of the Lagrangian associated with the selected constraints, omitting the terms for the aggregate procurement definition constraints (\ref{eqn: Indefine})--(\ref{eqn: PFRddefine}):
\vspace{-3mm}
\begin{multline*}
\lambda^E\!\left(P^{\rm D}\!-\!\sum_{i\in\mathcal{I}}P_i^{\rm E}\! -\!\sum_{j\in\mathcal{J}}P_j^{\rm E}\right)\! + \!\sum_{k\in\mathcal{I}\cup\mathcal{J}}\! \lambda^{k}\!\left(P_k^{\rm E}\! -\! \Delta P^{\rm L}\right)\\
+ \mu\!\left(\Delta P^{\rm L}\!-\!P^{\rm In}\right)
+ \delta\!\left(\Delta P^{\rm L}\! - \!P^{\rm PFR,d}\right) + \gamma_1\!\left(\frac{P^{\rm In}}{2\overline{\Delta f^{'}}}\! - \!\frac{P^{\rm PFR,r}}{T^{\rm PFR}}\right)\\
+\gamma_2\!\left( \frac{\Delta P^{\rm L}}{\sqrt{ \overline{\Delta f}_{\rm Nad}}}\right)\! -\! \gamma_3\!\left(\frac{P^{\rm In}}{2\overline{\Delta f^{'}}}\! + \!\frac{P^{\rm PFR,r}}{T^{\rm PFR}} \right)
\end{multline*}
\vspace{-1mm}

\begin{table}
    \centering
    \caption{Prices of energy and ancillary services}
    \renewcommand{\arraystretch}{0.9} 
    \vspace{-0.5mm}
    \begin{tabular}{l l}
        \hline
        \textbf{Service} & \textbf{Market Clearing Price} \\
        \hline
        Energy & 
        $\pi_{{\rm energy},k} = \lambda^{\rm E}, \quad \forall k \neq l$ \\[1mm]
        & $\pi_{{\rm energy},l} = \lambda^{\rm E} - \lambda^{l}$ \\[1mm]
        
        Inertia & 
        $\pi_{\rm inertia} = \mu + \dfrac{1}{2\overline{\Delta f^{'}}} \left(-\gamma_1 +\gamma_3 \right)$ \\[1mm]
        
        PFR (ramp) & 
        $\pi_{\rm PFR,r} = \dfrac{1}{T^{\rm PFR}} \left(\gamma_1+\gamma_3\right)$ \\[2mm]
        
        PFR (droop) & 
        $\pi_{\rm PFR,d} = \delta$\\
        \hline
    \end{tabular}\label{tab: prices}
    \vspace{-0.5mm}
\end{table}
\vspace{-0.25mm}
Applying the first-order Karush-Kuhn-Tucker (KKT) conditions yields the market clearing prices, as summarized in Table~\ref{tab: prices}. The resulting price formation represents key economic features of the co-optimized services. 

First, the energy price for the resource $l$ that sets the largest contingency, $\pi_{{\rm energy},l}$, is discounted by $\lambda^l$, effectively penalizing it for the additional security cost it imposes on the system. Furthermore, the AS prices reflect their marginal economic values in maintaining frequency security. The price of inertia, $\pi_{{\rm inertia},k}$, consists of $\mu$ and gammas, reflecting its contribution to both RoCoF and frequency nadir security. The price for PFR ramp is derived from the nadir constraint, as its primary value is in arresting the frequency drop. 

An analysis of the nadir-associated dual variables $\gamma_1$ and $\gamma_3$ provides further insight. These variables indicate the marginal value of each service's response capability, that is, the inertial response capability of $P^{\rm In}/(2\overline{\Delta f^{'}})$ and the PFR ramp rate $P^{\rm PFR,r}/T^{\rm PFR}$, in satisfying the nadir constraint. Specifically, $\gamma_3$ represents the common marginal value of these two capabilities in jointly meeting the nadir limit. It increases when the constraint is costly to satisfy, reflecting the growing economic value of these services. On the other hand, $\gamma_1$ captures the marginal value of the imbalance between the two capabilities. Its sign indicates which service is scheduled more heavily: a negative $\gamma_1$ implies that the PFR ramp capability is prioritized for maintaining the frequency nadir, whereas a positive $\gamma_1$ means that the inertia service capability is more favored. The absolute value of $\gamma_1$
represents the marginal cost of relaxing this imbalance. Finally, the droop PFR price, $\pi_{\rm PFR,d}$, is determined only by the QSS constraint ($\delta$).

\vspace{-4mm}
\subsection{Settlements}\label{subsec: Settlement}
\vspace{-0.5mm}

AS settlement is divided into two distinct payments: (1) service payments for reserved capacity commitments and (2) real-time deployment payments for actual service delivery. 

An AS provider $k$ receives a payment for each service equal to the product of the scheduled quantity and its price:
\vspace{-1mm}
\begin{align*}
    {\rm Payment}_{\rm inertia,k} &= \pi_{\rm inertia} \cdot P_k^{\rm In} \\
    {\rm Payment}_{\rm PFR,r,k} &= \pi_{\rm PFR,r} \cdot P_k^{\rm PFR,r}\\
    {\rm Payment}_{\rm PFR,d,k} &= \pi_{\rm PFR,d} \cdot P_k^{\rm PFR,d}.
\end{align*}
\vspace{-0.5mm}
These service payments compensate the providers for the opportunity costs of withholding power and energy capacity to provide each AS instead of participating in other markets. Additionally, resources are compensated for the actual energy delivered or absorbed during the event. Both PFR and inertia deployments are settled based on the energy provided at the RTP. However, because inertia deployment is bidirectional and closely coupled with PFR, service performance verification would rely on high-resolution frequency measurements. Let $P^{\rm IP,+}(t)$ and $P^{\rm IP,-}(t)$ denote the provider's positive and negative inertia provision, respectively. Then the inertia deployment payments for a provider $k$ are calculated as follows:
\vspace{-3mm}
\begin{align*}
    {\rm Payment}_{{\rm inertia},k}^{+} &= \pi_{\rm energy} \cdot \int P_k^{\rm IP,+}(t) dt \\
    {\rm Payment}_{{\rm inertia},k}^{-} &= \pi_{\rm energy} \cdot \int P_k^{\rm IP,-}(t) dt.
\end{align*}
Here, ${\rm Payment}_{\rm inertia}^{+}$ and ${\rm Payment}_{\rm inertia}^{-}$ represent the deployment payments for the positive and negative inertia provision over the dispatch interval. Under ideal conditions where a frequency event begins and fully recovers within a single dispatch interval, the net payment for inertial energy deployment is zero ignoring round-trip losses. However, for an IBR, round-trip efficiency losses result in a net energy consumption over the dispatch interval (i.e., it must release less energy than it absorbs). This net energy consumption exposes the IBR to RTP risks. As previously discussed in \cite{Garcia2025Co-OptimizationMarket}, since the net energy represents only seconds or minutes of inertia provision, the resulting price exposure is typically modest. Nevertheless, in some markets where the real-time market could be rerun following a contingency, IBRs may face substantially different prices between energy selling and purchasing periods, creating high-price exposure risks. VI providers are therefore expected to incorporate this price risk into their VI offers.

\vspace{-2mm}
\section{Case Studies} \label{Case Studies}
\vspace{-0.5mm}
To explore the validity of the proposed definition and mathematical formulations, simulations are conducted using a modified version of the IEEE 30 test system. 
\vspace{-4mm}
\subsection{Simulation Settings} \label{subsec: Settings}
\vspace{-1mm}
The test system consists of six generators and one IBR with detailed characteristics and cost functions for the generators presented in Table~\ref{tab: gens}. We consider the ED problem following UC with all generators assumed to be online. For the IBR, the inertia constant range is assumed to be 0--50s \cite{AEMO2023HornsdaleVMM}. The variable cost of energy service is set at \$3/MWh \cite{Kushwaha2023ASystems} to account for ESS degradation costs\footnote{Although variable operating and maintenance costs are often considered to be zero for ESSs, we assume a small amount of cost to prevent biased results, applying the energy cost to both energy dispatch and the loss term.}. The detailed simulation data is available at \url{https://github.com/Sojin-Park-io/Power-Oriented-Inertia-Definition}.
    
\begin{table}
    \centering
    \caption{Parameters of 6 generators}
    \vspace{-2mm}
    \setlength{\tabcolsep}{3.5pt}
    \renewcommand{\arraystretch}{0.9} 
    \begin{tabular}{c|c|c|c|c|c}
        \hline
        \textbf{Gen} & $\underline{P}_i$ (MW) & $\overline{P}_i$, $\overline{P}_i^{\rm Gov}$ (MW) &  $a_i$ (\$/MW\textsuperscript{2}) & $b_i$(\$/MW) & $c_i$ (\$) \\ \hline
        G1              & 24                         & 80                        & 0.02                  & 2            & 0     \\ \hline
        G2              & 24                         & 80                        & 0.0175                & 1.75         & 0     \\ \hline
        G3              & 15                         & 50                        & 0.0625                & 1            & 0     \\ \hline
        G4              & 16.5                         & 55                        & 0.0083                & 3.25         & 0     \\ \hline
        G5              & 9                         & 30                        & 0.025                 & 3            & 0     \\ \hline
        G6              & 12                          & 40                        & 0.025                 & 3            & 0     \\ \hline
    \end{tabular}
    \vspace{-5mm}
    \label{tab: gens}
\end{table}

\vspace{-4mm}
\subsection{Simulation Results} \label{subsec: Results}
\vspace{-1mm}
This section demonstrates the key features introduced by the proposed P-oriented definition, including the establishment of an inertia price in \$/MW and the explicit consideration of footroom for a bidirectional service. For this, we compare bidirectional inertia service with a positive response-only approach under two scenarios: a high-SI grid where all SGs have an inertia constant of 5.0s, and a low-SI grid with 1.0s.
\begin{figure}
  \centering
  \includegraphics[width=3.5in, height=1.7in]{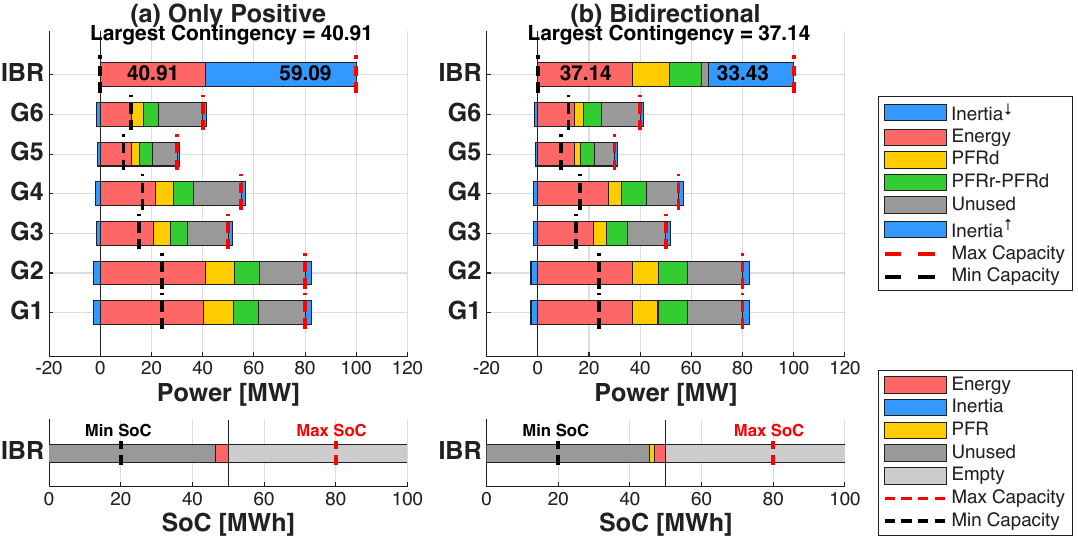}
  \vspace{-2mm}
  \caption{Power \& energy capacity allocation results from (a) only positive inertial response (b) bidirectional inertial response}
  \vspace{-2mm}
  \label{fig: sim1}
\end{figure}
\vspace{-1mm}
Table~\ref{tab: sim_1_1} presents the total operating cost and pricing results for both scenarios. In the high-SI grid, where inertia is abundant, the bidirectional requirement has no impact, resulting in identical outcomes for both cases as the footroom constraint (\ref{eqn: finalESSASLim2}) is non-binding. In the low-SI scenario, however, introducing the bidirectional service increases the total operating cost from \$546.92 to \$551.21 and results in an almost fivefold increase in the inertia price, from \$0.20/MW to \$0.96/MW. This increase in cost and price is not an inefficiency but rather a reflection of the true economic cost of ensuring service reliability by explicitly valuing the required footroom\footnote{Footroom is essential for ensuring that inertia providers can deliver a stable negative response. Without sufficient footroom during the FRP, energy absorption is constrained, which could lead to unstable frequency dynamics.}. 

An analysis of the dual variables in Table \ref{tab: sim_1_2} provides insight into the economic drivers behind this outcome. The zero values for $\mu$ and $\delta$ confirm that the RoCoF and QSS constraints are non-binding in these scenarios, indicating that the frequency nadir constraint is the primary factor setting the AS prices. Of the nadir-associated dual variables, the increase in $\gamma_3$, rising from 0.6665 to 2.0587, primarily accounts for the increase in the inertia price under the bidirectional service. By contrast, $\gamma_1$ remains positive, but declines from 0.2697 to 0.1465. This suggests that, while the inertia capability is still scheduled more heavily than PFR ramp capability, the marginal value of the imbalance diminishes. This reduction is explained by the fact that the PFR ramp is scheduled up to the maximum capability, fully exhausting the capabilities of both SGs and IBRs. As a result, the marginal cost of relaxing the imbalance actually decreases, as the scarcity-driven increase in PFR ramp value offsets the higher opportunity cost of inertia under bidirectional reservation. The decrease in $\gamma_1$ therefore reflects a shifting balance in relative cost-effectiveness between inertia and PFR.

\begin{table}
\setlength{\tabcolsep}{2.3pt}
\renewcommand{\arraystretch}{0.9} 
\centering
\caption{Comparison of total operating cost and service prices with bidirectional inertia service}
\vspace{-1mm}
\begin{tabular}{c|c|c|c|c|c|c}
    \hline
    \makecell{Inertia \\ Constants \\ (s)} & Case & 
    \makecell{Total \\ Cost (\$)} & 
    \makecell{System \\ Lambda \\ (\$/MWh)} & 
    \makecell{Inertia \\ (\$/MW)} & 
    \makecell{Ramp \\ PFR \\ (\$/MW)} & 
    \makecell{Droop \\ PFR \\ (\$/MW)} \\
    \hline
    \multirow{2}{*}{5.0} & Only Positive & 541.91 & 3.40 & 0.05 & 0.05 & 0.00 \\
                         & Bidirectional & 541.91 & 3.40 & 0.05 & 0.05 & 0.00 \\
    \hline
    \multirow{2}{*}{1.0} & Only Positive & 546.92 & 3.61 & 0.20 & 0.16 & 0.00 \\
                         & Bidirectional & 551.21 & 3.71 & 0.96 & 0.37 & 0.00 \\
    \hline
\end{tabular}
\label{tab: sim_1_1}
\vspace{-3mm}
\end{table}

\begin{table}
\setlength{\tabcolsep}{2.3pt}
\renewcommand{\arraystretch}{0.9} 
\centering
\caption{Comparison of dual variables with bidirectional inertia service}
\vspace{-1mm}
\begin{tabular}{c|c|c|c|c|c|c}
    \hline
    \makecell{Inertia \\ Constants \\ (s)} & Case & 
    \makecell{$\mu$} & 
    \makecell{$\gamma_1$} & 
    \makecell{$\gamma_2$} & 
    \makecell{$\gamma_3$} & 
    \makecell{$\delta$} \\
    \hline
    \multirow{2}{*}{5.0} & Only Positive & 0.0000 & 0.0979& 0.1695 & 0.1957 & 0.0000 \\
                         & Bidirectional & 0.0000 & 0.0979 & 0.1695 & 0.1958 & 0.0000\\
    \hline
    \multirow{2}{*}{1.0} & Only Positive & 0.0000 & \textbf{0.2697} & \textbf{0.6095} & \textbf{0.6665} & 0.0000 \\
                         & Bidirectional & 0.0000 & \textbf{0.1465} & \textbf{2.0535} & \textbf{2.0587} & 0.0000 \\
    \hline
\end{tabular}
\label{tab: sim_1_2}
\vspace{-3mm}
\end{table}

Fig.~\ref{fig: sim1} illustrates these insights through dispatch results for the low-SI scenario. The upper plots show the power capacity allocation for each resource, while the lower plots display the IBR's SoC. In Fig.~\ref{fig: sim1}(a), the positive-only case allows the IBR to maximize energy dispatch while providing substantial inertia capacity. Although a significant amount of inertia capacity is scheduled, its impact on the SoC is negligible given the short duration of inertial response relative to the MWh-scale energy capacity. 
Conversely, the bidirectional service case in Fig.~\ref{fig: sim1}(b) shows reduced commitments for both services. The scheduled inertia capacity of 33.43MW is equal to the energy dispatch of 37.14MW divided by the round-trip efficiency, indicating that constraint (\ref{eqn: finalESSASLim2}) becomes binding. Additionally, driven by the higher shadow price ($\gamma_2$), the largest contingency size decreases from 40.81 MW to 37.14 MW. Since increasing inertia capability would require expanding energy capacity and thus enlarging the contingency size, the IBR instead prioritizes securing substantially more PFR capability, demonstrating a more cost-effective approach.

\vspace{-3mm}
\section{Conclusion}
\vspace{-0.5mm}
This paper proposes the P-oriented perspective for inertia service that focuses on inertial response as the essence of inertia. Unlike the energy-based conventional concept of inertia, the proposed P-oriented definition quantifies inertia based on power delivery capabilities, ensuring functional consistency between SGs and IBRs.
This service framework makes inertia service compatible with existing ASs in terms of both units and conceptual meaning, facilitating co-optimization with other services. The proposed definition is expected to benefit future inertia market development by providing a standardized framework that accommodates diverse inertia providers while ensuring transparent and non-discriminatory compensation. However, the proposed definition does not incorporate broader IBR control technologies. Future research should develop metrics to quantify these advanced control features along with measurement technologies and design detailed settlement structures that appropriately value enhanced IBR contributions.
\vspace{-7mm}

\ifCLASSOPTIONcaptionsoff
  \newpage
\fi


\bibliographystyle{IEEEtran}
\bibliography{ref}

\end{document}